\documentstyle[preprint,aps]{revtex}
\begin{document}
\draft
\preprint{IC/93/329}
\preprint{ASITP-93-52}
\title {\large\bf ON THE CONDUCTANCE SUM RULE
 FOR THE HIERARCHICAL EDGE STATES
OF THE FRACTIONAL QUANTUM HALL EFFECT}
\author{\sf $^{a,b}$ Zhong-Shui Ma, $^b$ Yi-Xin Chen,
$^{a,c}$ Zhao-Bin Su}
\address{$^a$ Institute of Theoretical Physics, Academia Sinica,
P.O.Box 2735, Beijing 100080, China}
\address{$^b$ Zhejiang Institute of Modern Physics, Zhejiang University,
Hangzhou 310027, China}
\address{$^c$ International Center for Theoretical Physics,
P.O.Box 586, 34100 Trieste, Italy}
\date{\today}
\maketitle
\begin{abstract}
 The conductance sum rule for the hierarchical
edge channel currents of a Fractional Quantum Hall Effect state
is derived analytically within the Haldane-Halperin hierarchy scheme.
We provide also an intuitive interpretation for the hierarchical drift
velocities of the edge excitations.
\end{abstract}
\narrowtext

The fractional quantum Hall effect ( FQHE ) [1] has been an object of intense
experimental as well as theoretical investigations in the recent years [2].
One of the important features which has aroused a lot of interests is the
existence of the edge states for a finite FQH system [3-6]. It is generally
believed that, similar to that in the integer quantum Hall effect (IQHF)[7],
the transport of Hall current would rise on the edge and it can be in
principle detected by certain appropriately designed experiments.
But unlike the case of IQHE, in which there is a one-to-one correspondence
between the edge channels and bulk Landau levels [7], it has been
suggested [3,4]
that there exists a set of edge state branches hierarchically for
each FQHE state with filling factor $\nu$. Actually, the investigation
of the edge state
picture has been placed into the framework of Landau resistance formula
[3]. And furthermore, the low energy dynamics of the edge state has
been studied by means of the U(1) Kac-Moody algebra [5]
which is in fact equivalent to a chiral boson theory [5,6]. But such an
elegant effective field theoretical description is inferred from the general
principle but not derived from the microscopic Hamiltonian.

It is also believed that the Hall conductance of each hierarchical FQHE state
with filling $\nu$, i.e., $\nu e^{2}/ h$, is contributed by all its
associated branches of edge states. Therefore, there should be certain
sort of sum rule to describe such a fact hierarchically.  MacDonald
argued [4] that each branch of edge state associates with a
specifically assigned fractional charge $f_{i}e$, and the corresponding sum
rule has the form as $\sum_{i}f_{i}=\nu$.
On the other hand, Wen proposed a form of the sum rule as [5]
$ \sum_i ( {v_i \over | v_i |} ) q_i^2 = \nu e^2 $
where $v_{i}$ and $q_{i}$ are the velocities and ``optical" charges of the
fermions
in the $i$-$th$ branch of the edge waves.  In this paper, we apply the
Beenakker-MacDonald's arguments [3,4] for the conductance of the edge
current to the (constrained) Chern-Simons (C-S) field theory
approach for the finite FQH systems [8,9] in which the constraint for the
lowest Landau level (LLL) has been carefully considered;
derive an explicit expression for the branch conductance of each edge
channel; and show analytically the sum rule for the hierarchical
conductance by making use of the expressions for the drift velocities of
the edge waves derived in [9].  Meanwhile we provide further an intuitive
interpretation for the expressions of the hierarchical drift
velocities. Such a derivation, although going through
in the sense of weak coupling limit [9], provides one another exact relation
in the theoretical description for the FQHE. Therefore, the quantization
condition for the FQHE can be interpreted equivalently in terms of the
conductance for the transporting edge currents. It not only exhibits a
transparent understanding for the hierarchical structure of the edge
currents which
follows precisely the Haldane-Halperin [10,11] hierarchy scheme, but also
reveals a picture that, especially
in the context of a Chern-Simons' (C-S) field theory approach for the finite
FQH systems, the transport properties of the FQHE state are carried by the
compressible edge states while the bulk part keeps itself being an
incompressible liquid state.

To begin with, we would like to give a brief review of the relevant
results in our previous works [8,9].  In the (constrained)
C-S field theory approach
for the FQH system, the constraint for the LLL acquires a generic form which
can be transmitted from one hierarchical level to the next as
\begin{equation}
(-1)^n {\kappa_{n+1} \over 4 \pi} \nabla ^2 \ln \rho^{(n)} +
{1 \over 2 \pi m_{n+1}
\lambda ^2} - \rho^{(n)} - \kappa _{n+1} \rho ^{(n+1)} =0
\end{equation}
with $n=0,1,...$, where $\kappa_{n}$ and $m^{-1}_{n}$ are the
corresponding statistics index and
charge fraction of the fractionally charged quasi-particles at the $n$-$th$
hierarchical level respectively [12].  They have the inductive expressions as
\begin{equation}
\kappa _n = {1 \over \kappa_{n-1} + 2 p_{n-1} }
\end{equation}
\begin{equation}
m_n= {1 \over \kappa_1 \kappa_2 \cdot \cdot \cdot \kappa_n}
\end{equation}
with $m_{1}=1/\kappa_{1}=m$ being odd integers and $2p_n$'s being even
integers.   In
eq.(1), $\rho^{(n)}$ is the bulk density of the ``constituent"
quasi-particles of the $n$-$th$
hierarchical level while $\rho^{(n+1)}$ is the density of the ``excitational"
quasiparticles.   The latter could be interpreted as the ``vortex" density on
the background of ``condensed" constituent quasiparticle of the $n$-$th$
hierarchical level.  On the meanwhile, it gives rise itself also the density
of the constituent quasi-particles of the next
hierarchical level.  We separate these vortices into a ``surface" (boundary)
part and a bulk part with their densities $\rho^{(n+1)}_{surf}$ and
$\rho^{(n+1)}_{bulk}$ satisfying
$\rho_{surf}^{(n+1)} = (-1)^{n+1} (2 \pi)^{-1} \epsilon_{\alpha \beta}
\partial _\alpha \partial _\beta \theta^{(n+1)}_{surf}$
and $\rho^{(n+1)}_{bulk}= (-1)^{n+1} (2 \pi)^{-1} \epsilon_{\alpha \beta}
\partial _\alpha \partial _\beta \theta^{(n+1)}_{bulk}$
respectively, where $\theta^{(n+1)}_{surf}$ and $\theta^{(n+1)}_{bulk}$ are
the angle variables for the corresponding
vortices, $\alpha, \beta$ are the spatial index for a 2-D vector and
$\epsilon_{\alpha \beta}$ is the 2-D fundamental antisymmetric tensor.
Conceptually, the ``surface" vortices bears the physics of the
rippling of the boundary and will not contribute to the average vortex
density of the finite FQH system.  We may imagine also that the
locations of the ``surface" vortices forms a sort of boundary layer: the
edge of the finite system and $\theta_{surf}^{(n+1)}$ is the dynamical
variable of the
edge.  On the other hand, since only the bulk vortices
contribute to the average vortex density, therefore, for the corresponding
average densities $\bar \rho_{n}$ and  {$\bar \rho_{n+1}$}, we have
\begin{equation}
\bar \rho_n = {1 \over 2 \pi m_{n+1} \lambda^2} - \kappa_{n+1} \bar \rho
_{n+1}
\end{equation}
which follows from eq.(1) straightforwardly.
After applying a careful partial integration treatment to the actions of
the system, we have shown [8,9] that the action for the $n$-$th$ hierarchical
state can be splitted into two parts: a ``surface" part provides the action
for the $n$-$th$ branch of edge excitations while the remaining bulk part
is exactly the action for the $(n+1)$-$th$ hierarchical state.   In the
weak coupling limit [9], the actions for the $n$ branches of
edge excitations will decouple
from each other as well as from the bulk right at the $n$-$th$ hierarchical
filling.   Moreover, we derived analytically
the expressions of the drift velocities for all branches of edge
excitations from the ``surface" action as
\begin{equation}
v _D^{(n)} = { v_D \over 2 \pi m_{n+1} \lambda^2 \bar \rho ^{(n)}}
= {v_D \over 1 - 2 \pi \lambda^2 m_n \bar \rho ^{(n+1)} }
\end{equation}
with $n=0,1,2,...$ and $v_{D}=cE/B$.

To be specific, consider now a FQHE state of $(N-1)$-$th$ hierarchical level
for which we have its bulk quasi-particle density
$\rho^{(N)}_{bulk}=0$. In the sense of weak coupling limit,
there should be N branches of mutually independent edge states
associated with such a FQHE state, $n=0,1,...,N-1$.  It is natural to expect
that the Hall conductance
of the system is a simple sum of the branch conductance of each edge channel.
It is known that [4] the branch conductance $G_{n}$ for the $n$-$th$
edge channel has the expression as
\begin{equation}
G_n = q_n { \Delta I ^{(n)} \over \Delta \mu ^{(n)}}
\end{equation}
where $\Delta I^{(n)} $
is the variation of the Hall current induced by the variation of the chemical
potential contributed from the $n$-$th$ branch of edge states, $q_{n}$
is the fractional charge of the constituent quasi-particle of
$n$-$th$ hierarchical level and has the expression as
\begin{equation}
q_n= (-1)^{(n+1)} { e \over m_n}
\end{equation}
We notice that in the above equation we have taken $e > 0$, and the coefficient
$(-)^{n}$ is due to the reason as follows.  As it
can be easily seen in eq.(1), we have taken the convention that
the front signs of terms $\rho^{(n)}$ and $\rho^{(n+1)}$ are chosen to be the
same, i.e., the density $\rho^{(n+1)}$ is always counted as the hole-like
vortices with respect to the ``constituent quasi-particles" of the $n$-$th$
hierarchical level.  Therefore, in our convention, the fractional charge of
the quasi-particles should change its sign alternatively from each
hierarchical level
to the next.

We imagine now that the Hall current parallel to
the boundary is driven by a constant applied electric field normal to the
boundary. The induced variation of the Hall current for the $n$-$th$
branch of edge states  has the expression as [3]
\begin{equation}
\Delta I^{(n)} = q_n \Delta \rho^{(n)} v_n
\end{equation}
where $v_{n}$ is the moving velocity of the current carrying quasiparticles
parallel to the boundary and $\Delta\rho^{(n)}$ is the induced
variation of the linear density of the constituent quasiparticles of
the $n$-$th$ hierarchical level with $q_{n}$ being its fractional charge.
By definition, $\Delta\rho^{(n)}$ sums up all the variations of those
current carrying quasiparticles along the direction perpendicular to the
current flow (i.e., the boundary),
and depends subsequently only on the local position along the boundary.
Since the propagation velocity for any branch of the edge
excitations is constant, i.e., any sort of the ``surface" signals
should propagate with the same velocity, therefore, the drift
velocity for the transport current of the $n$-$th$ edge channel in
eq.(8), $v_{n}$, should take the same value as $v^{(n)}_{D}$. By
considering further a 2-D spatial integration
over the whole system, and utilizing
eqs.(1) and (4), we derive
\begin{equation}
 \int d^2 ( \rho^{(n)} - \bar \rho ^{(n)} ) = \oint_{\Gamma_n} d l n_\alpha
\Delta_\alpha \rho ^{(n)}
\end{equation}
with
\begin{equation}
\Delta_\alpha \rho^{(n)} = (-1)^n { \kappa _{n+1} \over 2 \pi }
(\epsilon_{\alpha \beta} \partial_\beta \theta^{(n+1)}_{surf}
+ { 1 \over 2} \partial_\alpha \ln \rho^{(n)})
\end{equation}
where $\Gamma_{n}$ is the boundary of the ensemble of the $n$-$th$
quasiparticles, i.e., the $n$-$th$ edge channel.
As can be easily seen in eq.(9), $n_{\alpha}\Delta_{\alpha}\rho^{(n)}$ is
the linear density which accumulates all the surplus ($n$-$th$ constituent)
quasiparticles apart from those belong to the the averaged bulk part along the
direction perpendicular to the boundary.  We may fix the meaning for the
``variation" in such a way that to identify the $\Delta\rho^{(n)}$ in eq.(8)
as
\begin{equation}
\Delta \rho^{(n)} = n_\alpha \Delta \rho_\alpha^{(n)}
\end{equation}

Consider next the integral form of the continuity
equation for the constituent quasiparticles of the $n$-$th$ hierarchical
level which is actually an identity in the first quantization representation.
We can express it in a form as [8,9]
\begin{equation}
\int d^2 x ( \rho ^{(n)}- \bar \rho^{(n)}) = - \oint_{\Gamma_n} d l
\rho ^{(n)} n_\alpha \delta r_\alpha^{(n)}
\end{equation}
where, as we like to emphasize, the left hand side
of the equation is the same as that of eq.(9) so that the variations introduced
here and in the following are really in consistency with those introduced
in eqs.(9)and (11).
In eq.(12), $\delta r_{\alpha}^{(n)}$ could be interpreted either as
the displacement for the $n$-$th$ quasiparticles passing back (accumulating)
and  forth (dissipating) through the boundary, or the ``rippling
displacement'' of the boundary deviating out- and inward along the
boundary.   Comparing eqs.(9), (11) and (12), we further have
\begin{equation}
\Delta \rho^{(n)} = - \rho^{(n)} n_\alpha \delta r_\alpha^{(n)}
\Big \vert_{\Gamma_n}
\end{equation}

Moreover, as a finite FQHE system, the ensemble of
the (condensed) $n$-$th$ quasiparticles are confined by certain envelop
potential.  And its chemical potential $\mu_{n}$ should be determined
in such a way that the Gibbs free energy is minimized consistently
with the spatial distribution of the $n$-$th$ quasiparticles.  As a
result, the local deviation of the applied electric field,
$e\varphi$, from the chemical potential at the surface boundary
is equal to the work done by those quasiparticles passed through the
boundary, or in other words, due to the local displacement of the
surface boundary from its equilibrium configuration.  Therefore,
to the first order of $\delta r_{\alpha}^{(n)}$, we have [8,9]
\begin{equation}
\Delta \mu ^{(n)} = - q_n E n_\alpha \delta r_\alpha^{(n)}
\end{equation}
where we assumed that
the applied electric field $E_{\alpha}$ being parallel to the normal of
the boundary: $E_{\alpha}=En_{\alpha}$.  Combining eqs.(13) and (14), we
have straightforwardly
\begin{equation}
{ \delta \mu^{(n)} \over \Delta \rho ^{(n)} } = { q_n E \over \rho^{(n)} }
\Big \vert _{\Gamma_n}
\end{equation}

Consequently, taking into all the above considerations, especially
eqs.(6), (8), (13) and (14), we derive the expressions for the branch
conductance as
\begin{equation}
G_n = q_n \bar \rho ^{(n)} v_D^{(n)} {1 \over E}
=(-)^{n+1}\kappa_{1}^{2}\kappa_{2}^{2}...\kappa_{n}^{2}\kappa_{n+1}
\frac{e^2}{h}
\end{equation}
with $n=0,1,2,...,N-1$. For the last equality of eq.(16), we have made
use of the expressions for the drift velocities
$v^{(n)}_{D}$ and the fractional charges $q_{n}$, eqs.(5) and (7)
respectively. Noticing further the identity
$\nu=\sum_{n=0}^{N-1}(-)^{n}\kappa_{1}^{2}\kappa_{2}^{2}...\kappa_{n}^{2}
\kappa_{n+1}$ [9], and then summing over all
the edge channels, we obtain straightforwardly the
conductance sum rule for the FQHE state with filling $\nu$ as
\begin{equation}
\displaystyle\sum_{n=0}^{N-1} G_n = - \nu { e^2 \over h}.
\end{equation}

For making the underlying physics clearer, we consider two special cases.
Consider first the $N=0$ case, i.e., the FQHE states of the lowest
hierarchical level characterized as $\rho^{(1)}_{bulk}=0$, and it associates
with only one edge channel.  The constituent quasiparticles are electrons
with $q_{0}=-e$ while the dynamics of the rippling boundary is carried by
the excitational quasiparticles, i.e., the ``surface" vortices upon the
electron liquid which is essentially the quasiparticles of the
hierarchical level $n=1$. It can be shown easily that the corresponding
drift velocity $v_{D}^{(0)}$ is equal exactly to $cE/B$ as we have
{${\bar \rho}^{(0))}$}=$(2\pi m\lambda^{2})^{-1}$ and $\nu=m^{-1}$.
For the $N=1$
case, it is the FQHE states of the next (to the lowest) hierarchical level
defined as $\rho^{(2)}_{bulk}=0$.  There should be two edge channels
associated with each FQHE states. The $n=0$ channel corresponds to the
constituent quasiparticles being electrons with a drift velocity of
$v^{(0)}_{D}=(m\nu)^{-1}cE/B$, while the $n=1$ channel corresponds to the
constituent quasiparticles being the $1/m$ fractionally charged bulk
vortex excitations (upon the condensed electron liquid)
 associated with a drift
velocity $v^{(1)}_{D}$ again equal to $cE/B$. A natural question
raised immediately: what is the physical meaning of the analytical
expressions eq.(5) for the hierarchical drift velocity, i.e., why they do
not always coincide the conventional drift velocity expression as $cE/B$.

In the classical electrodynamics, it is known that, driven by an
applied static electric field, charge particles in a strong magnetic field
would acquire a drift velocity as $c{\bf E}\times{\bf B}/B^{2}$ which is
independent on its charges and masses.  But the distinguished issue, here,
is that the quasiparticles of different hierarchical level will see a
different effective magnetic field.  For a finite FQH system, the bulk
action (in the first quantization representation) for the $n$-$th$
hierarchical level has the expression as [8,9]
\begin{equation}
I^{(n)} = {\bf j}^{(n)} \cdot {\bf A}^{(n)} - (q_n \varphi- \mu^{(n)})
\rho^{(n)} + (-1)^n {\kappa_n \over 4 \pi} \epsilon_{\alpha \beta}
A^{(n)}_\alpha \dot A^{(n)}_\beta
\end{equation}
with
\begin{equation}
j^{(n)} = - \sum_j c_j ^{(n)} \dot {\bf r }^{(n)}_j (t)
\delta ({\bf r} - {\bf r}_j^{(n)}(t))
\end{equation}
where {\bf $r$}$^{(n)}_{j}(t)$ is the trajectory of the $j$-$th$ quasiparticle
belong to the $n$-$th$ hierarchical level, $c_j^{(0)}\equiv -({e \over c})$ is
for the electrons while
$c^{(n)}_{j}=\pm 1$ with $n \geq 1$
is the vorticity for the corresponding quasiparticle, and {\bf $A$}$^{(n)}$'s
satisfy [8,9]
$\epsilon_{\alpha \beta} \partial_\alpha A^{(0)}_\beta = B$ for
$n=0$ case and
\begin{equation}
\epsilon_{\alpha \beta} \partial_\alpha A_\beta^{(n)} = (-1)^n 2 \pi \hbar
\rho^{(n-1)}
{}~~~~~~~~~~~~~~~~~~~n \geq 1
\end{equation}
In eq.(20), $\rho^{(n-1)}$ is the bulk density of the $(n-1)$-$th$
constituent quasiparticles. We can learn from the action expression
eq.(18), in cooperating with eqs.(19) and (20), that the quasiparticles
of $n$-$th$ hierarchical level with $n\geq1$, carrying a fractional charge
$q_{n}$, would couple
to the applied electric field in an usual way, but couple further to a
vector potential which is due to the condensation of the $(n-1)$-$th$
constituent quasiparticles and is usually different from that for
the applied magnetic field.  We may introduce an effective magnetic field
$B^{(n)}_{eff}$ as
\begin{equation}
B^{(n)}_{eff} = - { e \over q_n} \epsilon_{\alpha \beta} \partial_\alpha
A_\beta^{(n)}
\end{equation}
By taking further into account of eq.(20), we have immediately that
$B^{(0)}_{eff} = B$ for $n=0$ case and
\begin{equation}
B^{(n)}_{eff} = B ( 2 \pi \lambda^2 m_n \rho ^{(n-1)})
{}~~~~~~~~~~~~~~~~~~~~n\geq 1
\end{equation}
In eq.(22), since $m_{n}$ can be written as
$m_{n-1}/\kappa_{n}$ ( see eq.(3) ), the effective magnetic field of the
$n$-$th$ hierarchical level could equal to the applied magnetic field only
when the statistical flux density induced by
$\kappa_{n-1}\rho_{n-1}$ being canceled by the the corresponding magnetic
flux density $1/2\pi m_{n-1}\lambda^{2}$ at $(n-1)-th$ hierarchical
level.  Moreover, in our description [8,9], as mentioned above already,
the dynamics of the rippling boundary for the $n$-$th$ hierarchical level is
in fact described by the dynamical variable of its excitational
quasiparticles which is essentially the ``surface" part of the
quasiparticles of the next hierarchical level.  Therefore, at the $n$-$th$
edge channel, the $(n+1)$-$th$ ``surface" vortices will see an effective
magnetic field of $(n+1)$-$th$ hierarchical level $B^{(n+1)}_{eff}$.
Subsequently, we may expect simply by intuition that the drift velocity of
$n$-$th$ edge channel will have the expression as
\begin{equation}
v^{(n)}_D = c { E \over B^{(n+1)}_{eff}}
\end{equation}
Substituting eq.(22) into eq.(23), we find eq.(23) is exactly identical
to eq.(5) which was derived in [9].

As a final remark, it is interesting to note that the right hand side of
eq.(15) is always nonzero so that our approach is consistent
with the compressibleness of the edge states [3]. We understand such a
compressibleness along the boundary is essentially due to the rippling
degrees of freedom perpendicular to the boundary.  We therefore realize
such a picture for the FQHE states that the transport current is carried
by the compressible edge states while the bulk interior keeps in an
incompressible liquid state.

In summary, for each FQHE states, we derived the conductance sum rule for
the hierarchical edge channel currents in the framework of the
(constrained) C-S field theory approach for the FQHE systems [8,9], which
makes it possible that the quantization condition for the FQHE could be
equivalently interpreted in terms of transport edge currents.   But it has
a form different from that of MacDanold [4] as he has an interesting rule
of assigning the fractional charge of the associated edge states while we
follow regularly from Haldane-Halperin hierarchy scheme [10,11].
Moreover, we have not succeeded to find the details of Wen's sum rule [5],
so that we can hardly make a detailed comparison with his result.  We
hope our discussion might provide certain interesting insights for a
thorough understanding of the edge states of the FQHE states.

\bigskip
\bigskip
{\raggedright{\large \bf ACKNOWLEDGEMENT\\}}
\bigskip

One of the author (Z.B.S.) would like to thank Prof. D.H.Lee for a very
useful discussion.  He also likes to thank Prof. B.Sakita for his kind
advisement.  This work is partially supported by the NSFC, ITP-CAS and ICTP.

\end{document}